# Information and Order of Genomic Sequences within Chromosomes as Identified by Complexity Theory. An integrated methodology.

L.P. Karakatsanis[1], E.G. Pavlos[1], G. Tsoulouhas[1], G.L. Stamokostas[1], T.L. Mosbruger[2], J.L. Duke[2], G.P. Pavlos[1] and D.S. Monos[2]

[1]Research Team of Chaos and Complexity, Department of Environmental Engineering, Democritus University of Thrace, 67100 Xanthi, Greece

[2]Department of Pathology and Laboratory Medicine, The Children's Hospital of Philadelphia and Perelman School of Medicine, University of Pennsylvania, Philadelphia, Pennsylvania 19104, USA

Corresponding authors: L.P. Karakatsanis, DUTH, (karaka@env.duth.gr); D.S. Monos UPENN/CHOP (monosd@email.chop.edu)

**Abstract**

Complexity metrics and machine learning (ML) models have been utilized to analyze the lengths of segmental genomic entities like: exons, introns, intergenic and repeat/unique DNA sequences, in each of the 22 human chromosomes. The purpose of the study was to assess information and order that may be concealed within the size distribution of these sequences. For this purpose, we developed an innovative integrated methodology. Our analysis is based upon the reconstructed phase space theorem, the non-extensive statistical theory of Tsallis, ML techniques and a new technical index, integrating the generated information, which we introduce and named it Complexity Factor (COFA). The low-dimensional deterministic nonlinear chaotic and non-extensive statistical character of the DNA sequences was verified with strong multifractal characteristics and long-range correlations with significant variations per genomic entity and per chromosome. The results of the analysis reveal changes in complexity behavior per genomic entity and chromosome regarding the size distribution of individual genomic segment. The lengths of intron regions show greater complexity behavior in all metrics than the exonic ones, with longer range correlations, and stronger memory effects, for all chromosomes. We conclude from our analysis, that the size distribution of the genomic regions within chromosomes, are not random, but follow a specific pattern with characteristic features, that have been seen here through its complexity character, and it is part of the dynamics of the whole genome according to complexity theory. This picture of dynamics of the redundancy of information in DNA recognized from ML tools for clustering, classification and prediction. The results of the ML models identified the different degree of complexity profile of the distribution of the regions and their classification in original regions with high accuracy based on the set of complexity metrics. The variations of the dynamics correspond to specific biological characteristics. Finally, we used an unsupervised k-means clustering model and based on the COFA index, we discriminated sets of genomic entities per chromosomes which appears similar dynamics or dynamics which lives around a local center. We believed that these sets may contain common flows of information which are produced from fundamental laws and symmetries. The interdisciplinary approach of utilizing tools from complexity theory to ask questions regarding the segmental organization of the human genome can potentially reveal patterns hidden within the human DNA and therefore contribute to a better and more comprehensive understanding of potential interactions taking place among different regions.

**1. Introduction**

The DNA structure in the human genome reflects the entire evolution process from simple to high complex biological forms and organisms. Complexity theory indicates the existence of a strange and self-organizing dynamic process underlying the biological evolution process. As we have shown, in two previous studies (Pavlos et al., 2015; Karakatsanis et al., 2018) concerning the Major Histocompatibility Complex (MHC) DNA sequence, nothing in DNA structure is "junk" or useless. The DNA base sequence is constructed by nature as a long-range correlated self-organized system and emergent biological form through the co-evolution of biological and environmental subsystems. From mathematical point of view, nature realizes complex mathematical forms with spatiotemporal correlations. The nonlinear and strange dynamics describe the evolution of complex systems such as biological systems, as a nonlinear complex process including critical states and critical points, where the system can develop ordered states and forms throughout the development of long-range spatiotemporal correlations. This mathematically behavior of nature is self-consistently described by the non-equilibrium thermodynamics and the non-extensive statistical theory of Tsallis. Nature works thermodynamically for the development of non-equilibrium stationary thermodynamical states where the entropy function is maximized (Prigogine, 1978; Nicolis and Prigogine, 1989; Nicolis, 1993; Tsallis, 2009). The development of the complexity theory (Prigogine, 1978;1997), through the information theory can describe the redundancy of information in DNA.

That is according to the classical biological description, only 1.5% of the human DNA is translated into proteins. The rest was traditionally thought as "junk". In order to determine the role of the remaining part of the genome, many tries have been made, with the most notable one the Encyclopedia of DNA elements project (ENCODE data portal update, 2017). In order to shed light to the problem from different perspectives, different from the conventional ones, there is a significant increase in novel interdisciplinary approaches and methods. More specifically the last years, in order to understand better, the complex character of biological systems, such as the order of information in genome, the origins of autoimmune diseases, etc, a huge amount of big data are produced from many studies. Many researchers have developed computational methods to identify and characterize DNA motifs throughout the genome utilizing methods borrowed from the field of signal processing, information theory, non-linear dynamics and the non-extensive statistics-based methods (Kellis, 2014; Broomhead, 1986; Casdagli, 1989; Theiler, 1990.; Grassberger, 1991; Provenzale, 1992;. Lorenz, 1993; Klimontovich, 1994; Tsallis, 1988; 2002; 2004; Peng, 1992; Provata, 2011; Wu, 2014). These statistical metrics can be used to describe the dynamic characteristics and the structure and organization of the human genome. Many scientists (Voss, 1992; Li, 1992; Buldyrev, 1993; Stanley, 1994; 1999; Ossadnik, 1994; Mohanty, 2000) have introduced and studied the DNA random walk process as a basic physical process-model for the detection and understanding of the observed long-range correlations of nucleotides in DNA sequences. According to Voss (Voss, 1992) there is no significant difference between long-range correlation properties of coding and noncoding sequences and in contrast, other scientists (Peng, 1992; Li, 1992; Buldyrev, 1993; 1995; Grosberg, 1993) found that the noncoding sequences appeared long-range correlations while coding sequences do not. This could mean that the dynamics which produced the spatial information of the DNA can be characterized by strange dynamics such as strange attractors, islands, multifractal behavior in the reconstructed phase space. Some significant theories from the account of the statistical physics, like self-organized criticality (SOC), strange dynamics, non-extensive statistical mechanics of Tsallis, fractional dynamics etc. have been proposed to interpret the development of long-range correlations of the DNA sequences. Recent research utilizing these methods has sought to characterize the structural characteristics of DNA (Provata, 2011; Woods, 2016; Namazi, 2015; Pavlos, 2015; Vinga 2007; Oikonomou, 2006; 2008; Karakatsanis, 2018).

More specifically, Melnik and Usatenko (Melnik, 2014), using an additive Markov chain approach, analyzed DNA molecules of different organisms and they estimated the differential entropy for the biological classification of these organisms. Similarly, Papapetrou and Kugiumtzis (Papapetrou, 2014;2019) studied DNA sequences, through the estimation of the Markov chain orders and Tsallis conditional mutual information. The results showed a different long memory structure in their DNA samples (coding and non-coding). In another study, Provata et al. (Provata, 2014) analyzed the evolutionary tree of higher eukaryotes, amoebae, unicellular eukaryotes and bacteria with complexity tools to estimate the conditional probability, the fluxes, the block entropy and the exit distance distributions. The study detected the changes in the statistical and complexity measures of the five organisms and proposed these measures as alternative methods for organism classification. Wu (Wu, 2014) studied the Synechocystis sp. PCC6803 genome by using the recurrence plot method and the technique of phase space reconstruction. This analysis revealed periodic and non-periodic correlations structures in the DNA sequences. Costa (Costa, 2019) and Makado (Makado, 2019) used tools from Kaniadakis statistics, fractal and information theory to uncover the order information of the Homo sapiens DNA chromosomes.

The last years the next generation sequencing (NGS) technologies produced a great set of big data with high accuracy and resolution, which permits to display changes at different biological scales. Thus, there is a need for an interdisciplinary study and this development is going to increase in the near future. Our work will facilitate that development, through our integrated methodology, which is very general, and through our future endeavors is going to be embedded in a website, for every researcher to input their genome sequence of interest for analysis and get their complexity metrics. The produced data from NGS have given to researchers the opportunity to develop machine learning techniques to analyze their properties, with models for clustering, pattern recognition, classification and prediction with supervised and unsupervised learning (Apostolou, 2019; Washburn, 2019; Varma, 2019; Frey, 2019; Manogaran, 2018). The using of methods based on statistics and machine learning algorithms have many common but also separate routes with respective disadvantages and advantages. The limit between statistics and machine learning is often no visible (Xu, 2019; Bzdok, 2018). For a review of the machine learning applications in genetics and genomics, you can see at ref. (Maxwell, 2015). The goal of all these methods of analysis should be the deep understanding of the biological system.

In this work, we expand upon our previous studies (Pavlos, 2015; Karakatsanis, 2018), were we measured the dynamical and the non-extensive statistical characteristics in the DNA sequence of the whole MHC as a single unit and in the exonic, intronic, intergenic sequences of the MHC as separate and independent entities, and we seek to identify order information included in the whole genome and their possible interactive relationships focus in regions such as: exons, introns, repeat and unique. Our analyses are based upon the reconstructed phase space theorem (Takens, 1981; Theiler, 1990), the non-extensive statistical theory of Tsallis (Tsallis, 1988; 2004; 2009), machine learning techniques and we introduce a new technical index, which we call it complexity factor (COFA), as a more suitable one to our analysis.

Our paper is organized as follows. In Sec. II we present theoretical highlights of the complexity theory, the data and the methodology we used for our analysis. In Sec. III we give a summary of our results, including the metrics we used from complexity theory, the new complexity factor we introduced (COFA), and the machine learning algorithms for classification, clustering and prediction. In Sec. IV we continue with the discussion. And finally, we leave for the appendix the technical theoretical framework of our analysis.

## 2. Data generation and analysis

This section has been organized to: first, briefly present some useful theoretical concepts obtained by the modern complexity theory development, second, we describe the source of DNA data, and then we present the integrated methodology of the analysis based on the more extended theoretical description of the appendix.

### 2.1 Theoretical high lights of complexity theory

The Complexity theory can give useful quantitative parameters for the description of DNA structure. Such quantities are the Tsallis q-triplet ($q_{sen}, q_{rel}, q_{stat}$), the Correlation dimensions, the Hurst exponent, the Lyapunov exponents, etc. According to Prigogine and Nicolis, far from equilibrium, nature can produce spatiotemporal self-organized forms through the extended probabilistic dynamics of correlations in agreement with the extended entropy principle (Prigogine, 1978; Nicolis and Prigogine, 1989; Nicolis, 1993; Prigogine 1997; Davies 2004). Far from equilibrium the entropy principle creates long-range correlations, as nature works to maximize the non-equilibrium entropy (Tsallis) function. The entropy principle for the far from equilibrium and open physical systems, as the biological systems are, leads to the creation of dissipative structures and self-organized multi-level and multi-scale long-range correlated physical forms. The maximization of Tsallis q-entropy can explain the formation of DNA structure as a non-equilibrium intermittent turbulence structure and a multiplicative self-organization process (Pavlos et al., 2015). From this point of view, the DNA structure is a constructed multifractal system of the four DNA bases (A, C, G, T) with high information redundancy. This mean that nothing in the DNA structuring is useless "junk", but every region of DNA sequence is useful, but not fully understood.

The underlying intermittent DNA turbulence which constructs the DNA sequence and the chromosomic high ordered system is mirrored in the well-known q-triplet of non-extensive statistical theory of Tsallis including three characteristic parameters ($q_{sen}, q_{rel}, q_{stat}$). The $q_{sen}$ parameter, describes the entropy production and the information redundancy, as the DNA sequence is constructed by the underlying DNA turbulence process, as multifractal DNA structure. The $q_{rel}$ parameter describes the relaxation process of the DNA turbulence system to the meta-equilibrium stationary state of DNA structure, where the q-entropy ($S_q$) of Tsallis statistics is maximized. The meta-equilibrium state with maximized entropy function corresponds to the chromosomic DNA system. The $q_{stat}$ parameter describes the statistical probability distribution function of the DNA complex or random structure at the DNA turbulent stationary state. The DNA turbulence system can be described dynamically as an anomalous random walk process creating the DNA bases series. This dynamic can include critical points where the DNA turbulence dynamics can change. This constructive biological evolutionary phase transition process can develop the entire self-organized multifractal dynamical system. The variations of the Tsallis q-triplet along the DNA sequence is the quantitative manifestation of the biological evolution process throughout the constructive scenario of critical DNA turbulent phase transition processes.

The multifractal character of this biological evolutionary process is mirrored at the evolution of the Hurst exponent along the DNA sequence. As the Hurst exponent changes along the DNA sequence it mirrors the degree of the multifractal character along the DNA structure.

The DNA structure can be explained as the dynamical evolution of the biological complex system in the underlined natural state space to the DNA turbulence dynamics. This state space can be reconstructed by numbering the DNA sequence, supposing that the constructed DNA sequence corresponds also to the temporal aspect of the DNA sequence. This means that the natural numbering of DNA bases corresponds to the temporal evolution of DNA structuring. This permit us, to use the embedding theory of Takens (Takens, 1981) for the multidimensional reconstruction of the state space underlying to the DNA dynamical process. This reconstructed state space describes the entire temporal biological evolution physical process. The DNA sequence is the one-dimensional time projection of the DNA Turbulence phase space in the form of the DNA "time series". The DNA reconstructed state space can explain the multifractal structuring of DNA sequence and mirrors the sequence of evolutionary phase transition biological process as the topological phase transition process of the biological dynamical state space topology. The DNA correlation dimension can be estimated in the reconstructed DNA state space, as well as, other useful geometrical and dynamical parameters of the biological evolution, can also be estimated (Pavlos et al, 2015; Karakatsanis, 2018).

After all, the DNA structure mirrors also the multifractal topology of the underlying DNA turbulence state space, as well as, the critical DNA phase transition process during the biological evolution of species related to the DNA construction through non-linear strange dynamics. Moreover, the self-consistently to the DNA strange dynamics maximization of Tsallis q-entropy, structures the DNA state space multifractal topology.

According to these theoretical concepts, in this study we use the sizes of DNA regions for the reconstruction of DNA state space according to Takens embedding theory. All the estimated parameters are related with the fractal topology of DNA state space. The changes of the complexity metrics correspond to the evolutionary topological phase transition process of the DNA state space, as well as of the underlying intermittent turbulence process of the chromosomic dynamical self-organization.

**2.2 Source of DNA Data**

The Genomic compartments we used in this study and the Gene definitions are taken from National Center for Biotechnology Information (NCBI) RefSeq. We used both the curated and predicted definitions: *http://genome.ucsc.edu/blog/the-new-ncbi-refseq-tracks-and-you/*) as described below:

a) Genic: RefSeq the lengths of all the genic regions were merged to one sequence of lengths per chromosome. (to account for alternative transcripts and overlapping genes). Intervals in this file represent sections of the region containing genes.
b) Intergenic: Intervals between the genic intervals. The lengths of all the intergenic regions were merged to one sequence of lengths per chromosome.
c) Exonic: RefSeq the lengths of all the exonic intervals were merged to one sequence of lengths per chromosome to account for alternative transcripts and overlapping genes. Intervals in this file represent sections of the region containing exons.
d) Intronic: Intervals within genic intervals and between exonic intervals. The lengths of all the intronic intervals were merged to one sequence of lengths per chromosome.
e) Repeat individual: Individual repeat intervals. This file contains the repeat name, repeat class and repeat family. The repeat class is the broadest group and matches what we viewed in integrative genomics viewer (IGV). The lengths of all the repeat individual intervals were merged to one sequence of lengths per chromosome.
f) Repeat merge: Merged repeat intervals.
g) Nonrepeat: Intervals between the repeat intervals.

In the following, we calculated the lengths of genome entities (genes, exonic, intronic, intergenic, repeat individual, repeat merge, unique).

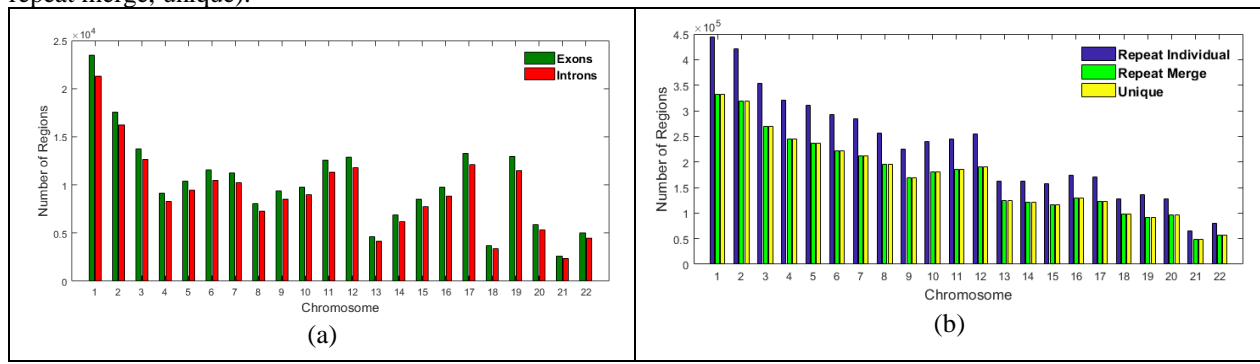

(a)  (b)

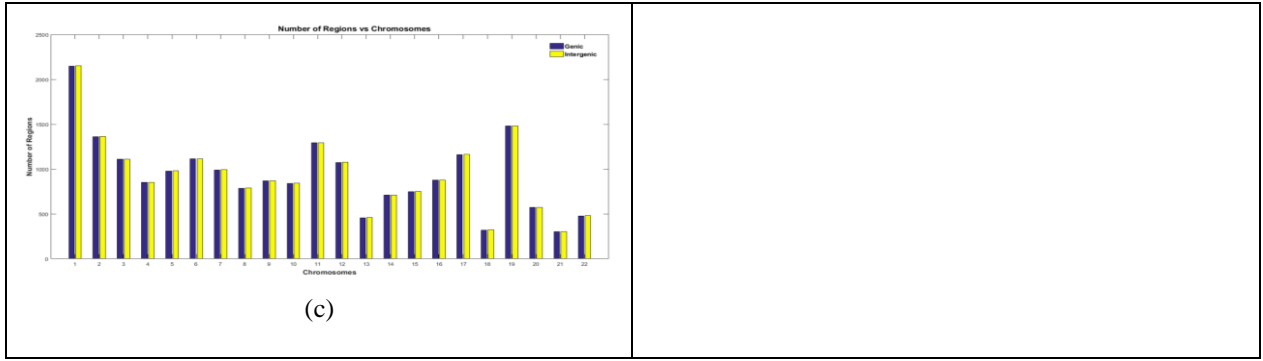

*Figure 1:* *The set of data for the analysis in 22 Chromosomes and 8 regions:* (**a**) *Exons, Introns,* (**b**) *Repeats, Unique,* (**c**) *Genic, Intergenic.*

In Figure 1 (a-c) we show the set of data, which includes 8 regions per chromosome (genic, intergenic, exons, introns, repeat individual, repeat merge, unique). The dataset equates 8 regions by 22 Chromosomes = 176 raw data. Finally, we analyzed 6 regions (exons, introns, repeat individual, repeat merge, unique) by 22 Chromosomes = 132 raw data. The regions of genic and intergenic they do not have enough number of points to satisfy the statistics in the reconstruction state space.

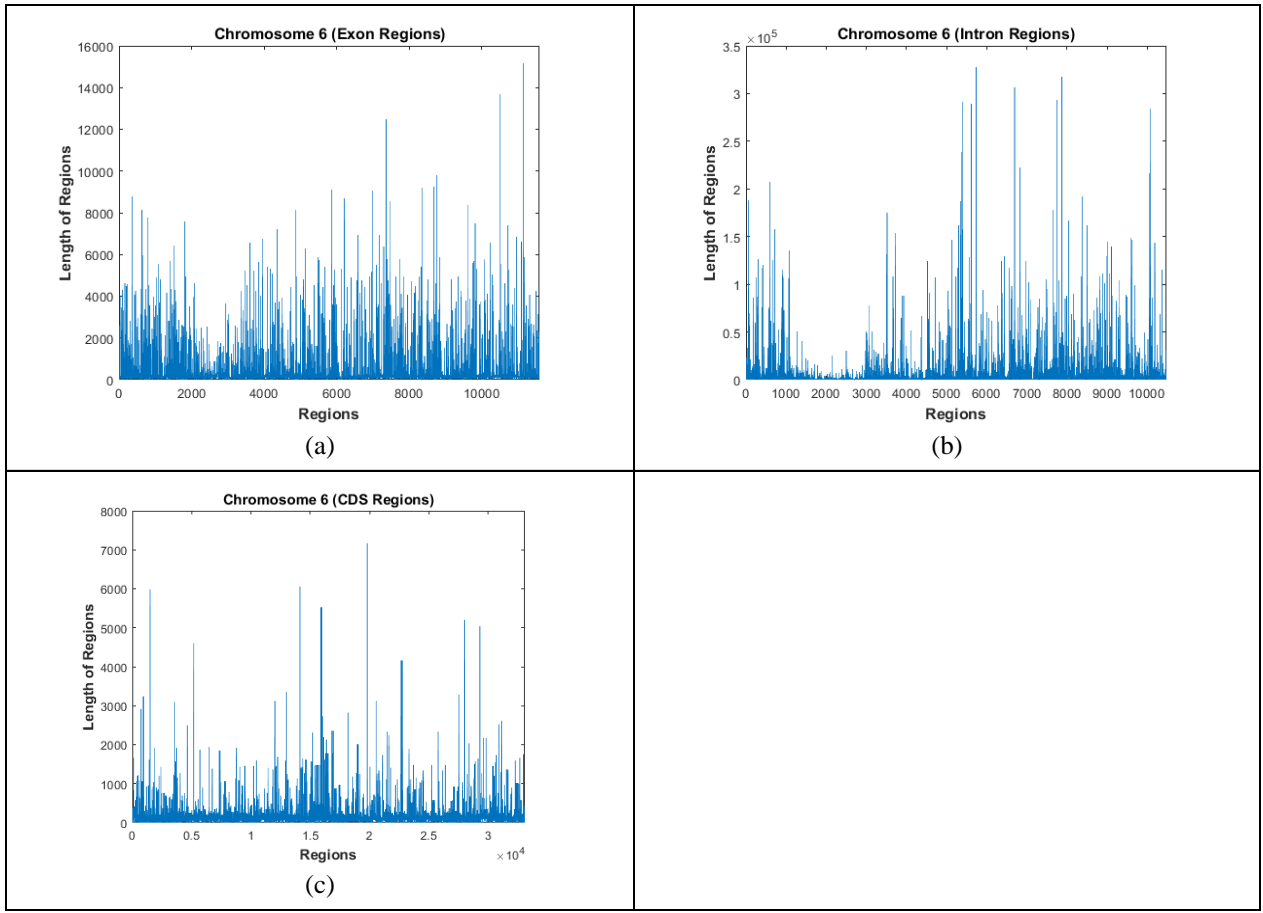

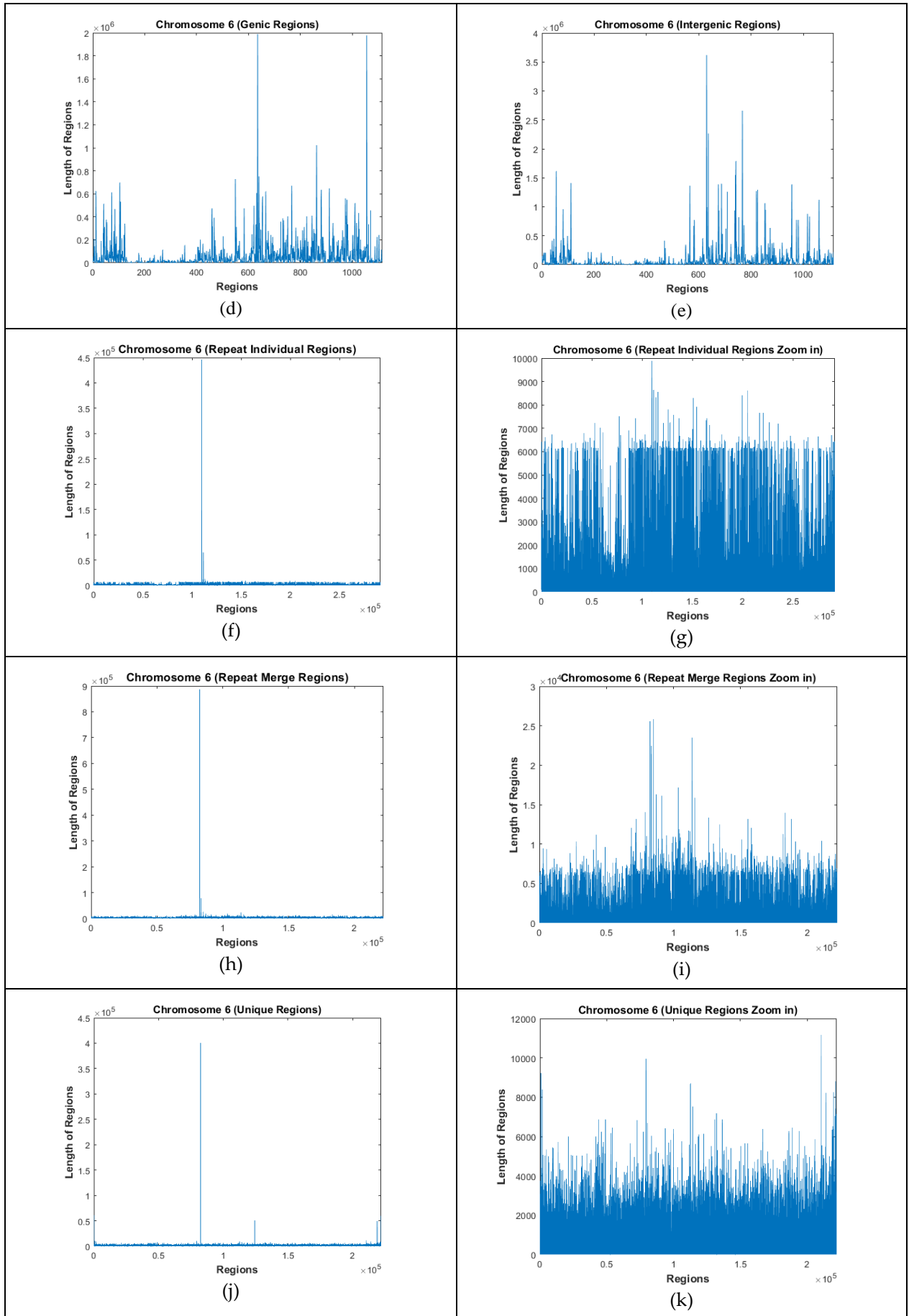

*Figure 2:* A sample of raw-data to all regions from chromosome 6 (*a-k*) We see clearly here that the lengths of regions have a fractal shape, indicating a complex behavior.

In Figure 2, we present a sample of raw data (space series) for all regions from Chromosome 6. For example, each point on axis $x$ correspond to the $i^{th}$ exon, intron, etc., while each point on axis $y$ correspond to the length of $i^{th}$ exon, intron, etc.

**2.3 Methodology of data analysis**

The methodology of the analysis of data are supported from metrics in physical and phase space. In order to unravel the symmetries and the order of information on the distribution of the lengths of the regions in all genome we used complexity theory tools for data analysis such as: a) q-triplet estimation, b) estimation of correlation dimension and c) estimation of hurst exponent on these data. Analytical description of the complexity metrics can be found in Appendix A. A new technical factor, the complexity factor (COFA) and tools from machine learning (ML) algorithms are used to describe better the variation of the metrics between genomic entities, with ultimate goal the deep understanding of the DNA system.

The dynamics of the DNA system in the phase space determines in the physical space the position of the fundamental four bases in the DNA chains in all genome entities. We understand that these positions included the necessary information for the following functions of DNA chains with an extended conclusion that the distribution of the genome entities are not random, but it is a part of the dynamics. The information we get from a measure quantity from the physical space, is a part of the projection of the dynamics which produced this physical and measure quantity. The complexity metrics we used in the analysis reflected every time part of the dynamics in the physical or phase space. The statistics of the information and the dynamics are inextricably linked in a continuous interaction from physical space to phase space and vice versa. The dynamics produces in the phase space objects with strange geometry like strange attractors, islands, long range correlations, diffusion, multifractal behavior etc. Staying in that line of thinking, we supposed that the variations of the metrics in different entities of the whole genome corresponds to changes of the strange dynamics in the phase space, marking entities like regions, words, etc in the genome with the scope to input such information in supervised or unsupervised ML models and help us to uncover patterns and symmetries of information in the whole genome.

**2.4 Complexity Factor (COFA)**

We introduce a new technical factor which characterize the degree of complexity in the phase space, taking in the account the set of complexity metrics which we used in the analysis:

$$COFA = \left(\sqrt{(q_{stat})^2 + (q_{rel})^2 + (q_{stat})^2}\right) h / D_2 \qquad (1)$$

where $q_{stat}, q_{rel}, q_{sen}$ are the $q$-triplet indices from Tsallis non-extensive statistics, $h$ is the hurst exponent and $D_2$ is the correlation dimension.

The scale of the factor appears the degree of complexity in the phase space in the metric of the Euclidean space. For a pure Gaussian (random) signal the Euclidean distance of the q-triplet equals to 1, the $h$ equals to 0.5 and for embedding dimension $m = 10$ the estimation of the $D_2$ gives a value $\cong 10$, so the COFA estimation is: $COFA = \frac{1 \times 0.5}{10} = 0.087$. On the following table we show the estimation of COFA for a various known models:

**Table 1. COFA for known models**

| MODELS | HURST | D2 (m=10) | qstat | qrel | qsen | EYKL | LINEAR COFA |
|---|---|---|---|---|---|---|---|
| Gaussian (theoretical) | 0.5 | 10 | 1 | 1 | 1 | 1.732 | 0.087 |
| White noise | 0.491 | 9.18 | 1.00 | 1.00 | 1.000 | 1.732 | 0.093 |
| Henon map | 0.415 | 1.26 | 1.00 | 1.00 | 1.000 | 1.732 | 0.570 |
| Logistic map | 0.466 | 0.54 | 1.65 | 2.25 | 0.240 | 2.800 | 2.417 |
| xLorenz | 0.621 | 2.12 | 0.93 | 1.15 | -1.236 | 1.926 | 0.564 |

The complexity factor creates a metric which characterizing the amount of the strange dynamics in a geometrical Euclidean space and it can be used as an external classifier to the ML modelling. The complexity factor is a linear transformation of the complexity metrics which we used. In future studies a non-linear transformation of the factor will be presented as well.

**2.5. An integrated analysis method of DNA entities**

In Figure 3 the general flow-chart of the analysis method of DNA data (arithmetic or text) is shown. This general method can be an innovative view of the DNA entities in all genome based on the strange dynamics with goal to reveal new symmetries and rules on this information.

We presented in algorithmic steps the whole methodology:

a) We prepare the arithmetic or text data from DNA system. If the data are text (independence bases, words, etc) we apply specific routines: like the distance a basis to the next similar basis or other methods, to transform the text data in arithmetic data, else we go to the next algorithmic step.
b) We apply on arithmetic data the complexity metrics like: hurst exponent, q-triplet of Tsallis, correlation dimension, etc (other complexity metrics).
c) We produce the table of results for the whole data set. In parallel mode, in this step we estimate the complexity factor (COFA), which it will be used for external classifier on ML models.
d) In this step we choose the attributes (hurst, q-triplet, etc), which will be used for a various ML model.
e) We apply ML models for classification, clustering and prediction based on the external classifier COFA.
f) We produce the table of accuracy from the previous step. If the accuracy is not acceptable, we return to step d) and we repeat the procedure until the accuracy be acceptable.
g) We present the final results from ML models and we extract the final symmetries and laws of information from the analysis of the whole data set.

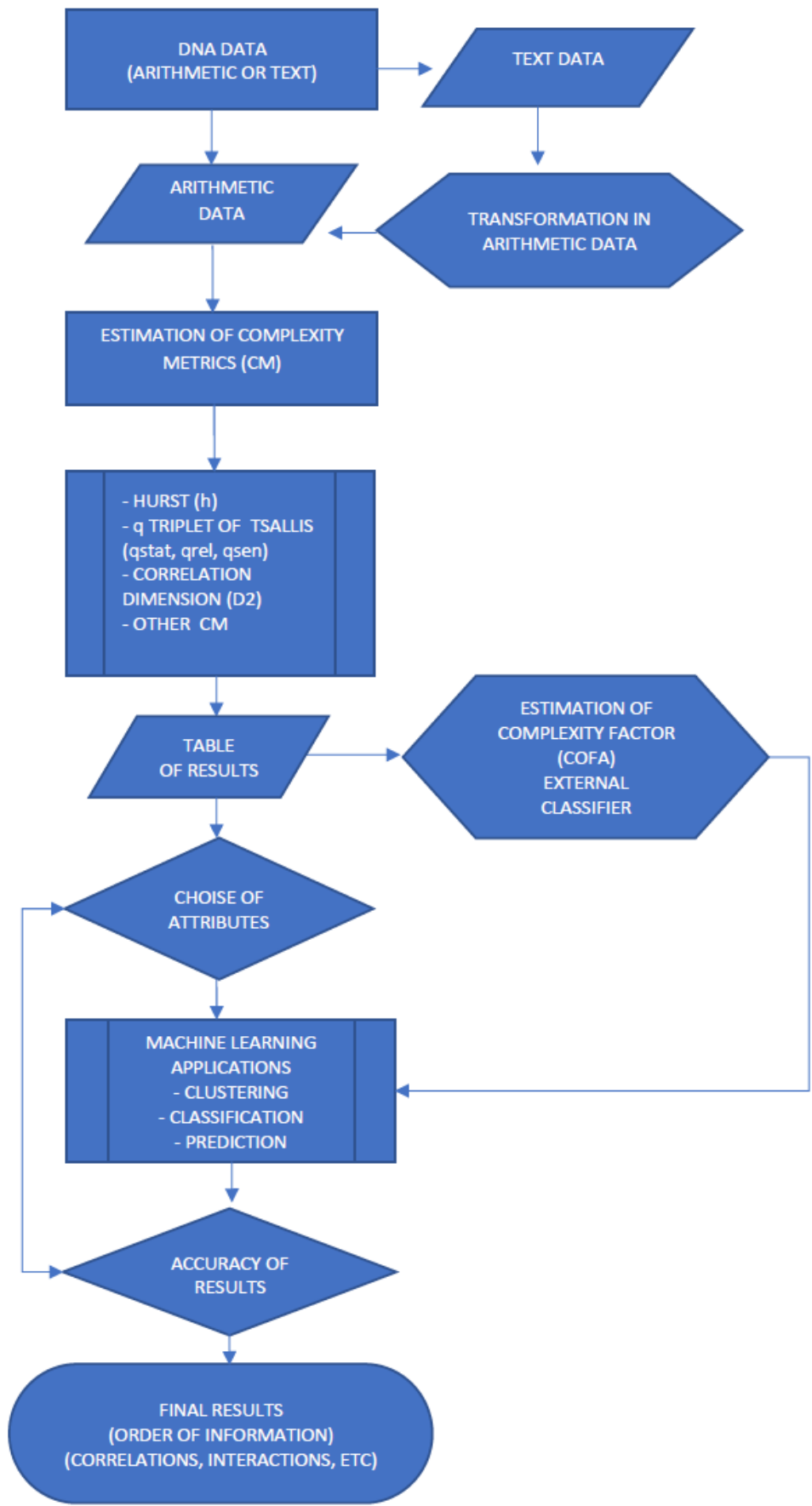

***Figure 3:*** *The flow-chart diagram of the method of analysis.*

## 3. Summary of the Results

### 3.1 Metrics of Complexity Theory

In this section the results from the estimation of complexity metrics in the distribution of the entities of genome are presented.

#### *3.1.1 Hurst exponent*

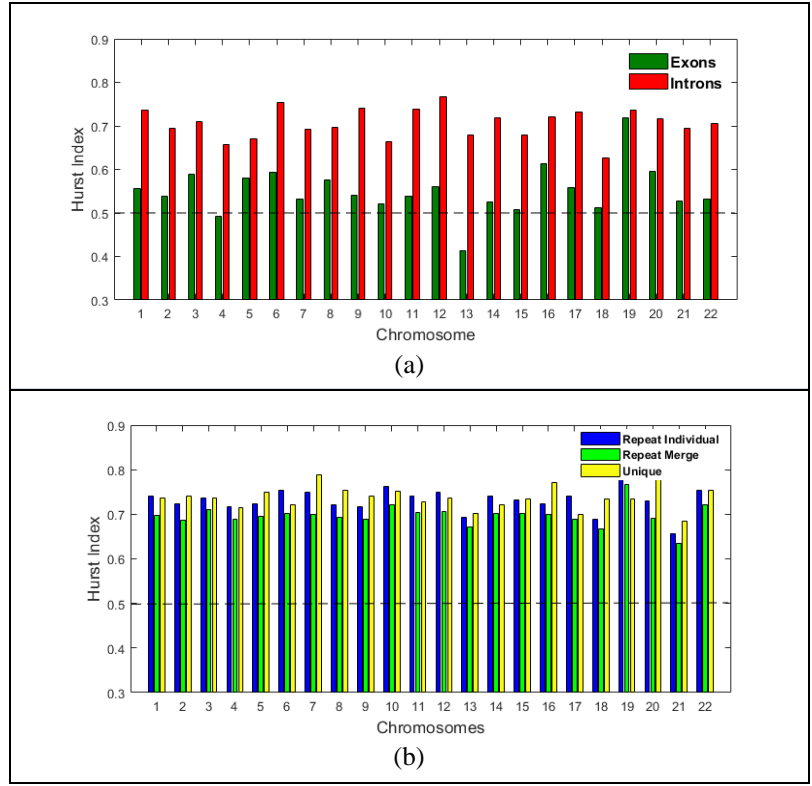

*Figure 4: The estimation of Hurst exponent per chromosome and genomic entity. (a) Exons, Introns, (b) Repeats, Unique.*

The Hurst exponent was estimated, for all genomic entities and for each Chromosome. Figure 4 (a-c) presents the estimated values of the Hurst exponent. The dashed line at 0.5, corresponds to a normal diffusion random walk process. As can be observed in Figure 3a, the Hurst exponent for intron data are much higher than 0.5 for all Chromosomes and related with persistent (super-diffusion) random walk process. For the exon data, the Hurst exponent is higher than 0.5 and related with persistent (super-diffusion) random walk process for all Chromosomes except Chromosome 13, which is lower than 0.5 and related with anti-persistent (sub-diffusion) random walk process. These findings reflect, the degree of the multifractal character and the existence of different scaling along the distribution of the DNA entities. For Chromosomes 4, 15 and 18, the Hurst exponent is almost equal to 0.5 and related with normal diffusion random walk process. This means that the profile is mono-fractal, and not permit the existence of different scaling in the data. Similarly, in Figure 3b, for the Repeat Individual, Repeat Merge and Unique data, the Hurst exponent is much higher than 0.5 for all Chromosomes and related with persistent (super-diffusion) random walk process.

#### *3.1.2 q-triplet of Tsallis statistics ($q_{stat}$, $q_{rel}$, $q_{sen}$)*

In Figures 5-7, we present the estimation of Tsallis q-triplet for all genomic entities and for all Chromosomes. Specifically, in Figure 5 we present the estimation of $q_{stat}$ index, in Figure 6 the estimation of $q_{rel}$ index and finally in Figure 7 the estimation of $q_{sen}$ index.

Concerning the $q_{stat}$ index (Figure 5), as one can see, the value in all chromosomes in all genomic entities is higher than 1 and suggests the presence of long-range correlations, a distinctive property of open non-equilibrium systems, with underlying dynamics characterized by non-Gaussian (q-Gaussian) distributions. The variations of the Tsallis $q_{stat}$ along the sizes of DNA entities is the quantitative manifestation of the biological evolution process throughout the constructive scenario of critical DNA turbulent phase transition processes. The development of long-range correlations means that the sizes of regions that are furthest between them are governed by fundamental rules on their size. Specifically, for the exonic genomic entity the index takes values mainly between 1 and 1.5, while for intronic one take values between 1.5 and 3. This mean, that the non-extensive character of the dynamics is much higher in introns than the exons and presents stronger long-range correlations in introns. Moreover, we observe a significant differentiation of the $q_{stat}$ index between chromosomes in both exonic and intronic genomic entities, which means a significant differentiation of the non-extensive character of the dynamics between chromosomes in the same genomic entity (Figure 5a). Similar in Figure 5b, the value of $q_{stat}$ index is higher for the repeat genomic entities than the unique one, which means that in the repeat the non-extensivity is higher than the unique region. However, between chromosomes within the same genomic entity there is not observed a significant differentiation.

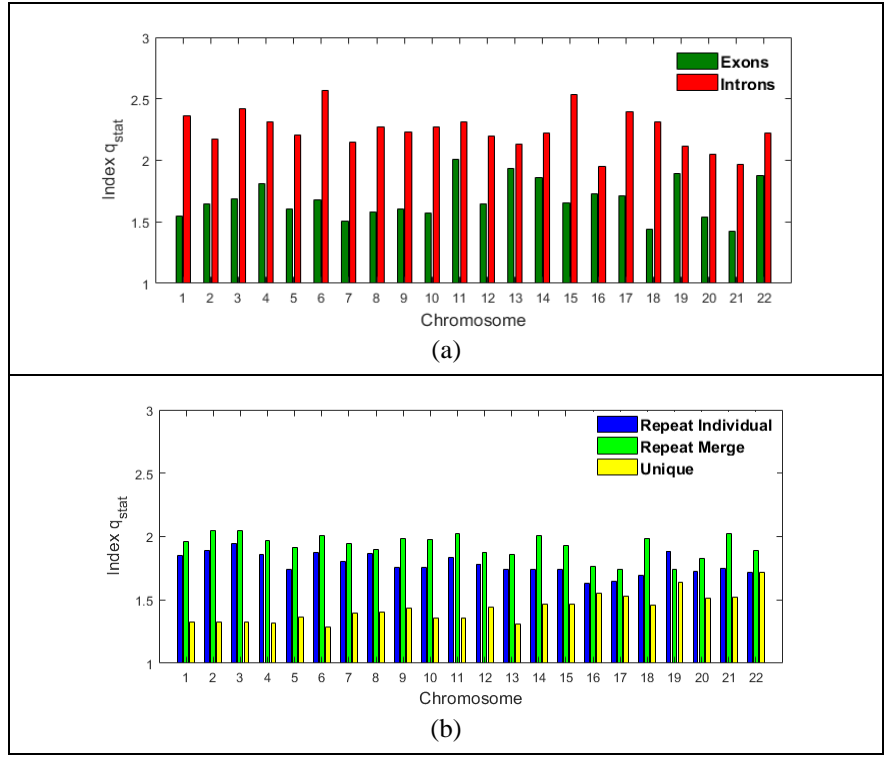

*Figure 5: The estimation of $q_{stat}$ index per chromosome and genomic entity: (a) Exons, Introns, (b) Repeats, Unique.*

As concern the $q_{rel}$ index (Figure 6), there is a significant differentiation between Exons and Introns (Figure 6a). As we observe, for all chromosomes the $q_{rel}$ index is higher for the introns than the exons. This reveals a non-Gaussian ($q_{rel} > 1$) relaxation process of the system to its nonequilibrium steady states (NESS) for the data in introns, while for the signals in exons reveals a near-Gaussian ($q_{rel} \approx 1$) or Gaussian ($q_{rel} = 1$) relaxation process of the system to its NESS. The results of the $q_{rel}$ in these regions, suggest that the distribution of the sizes may reach a new metastable state with different time (space) profiles. Clearly though while all regions include information, they are of a complex character, such that there are differences in the degree of complexity and therefore this complexity impacts the time (space) they take to transition to a new state of equilibrium upon being disturbed. In Figure 6b, we observe that in both repeat and unique regions the $q_{rel}$ index is different than 1 and this reveals a non-Gaussian relaxation process of the system to its NESS. However, in certain chromosomes the $q_{rel}$ index for one genomic entity is different than other(s) which means that in those cases the non-Gaussian relaxation process is stronger. Moreover, we observe differentiations of relaxation process between chromosomes within the same genomic entity.

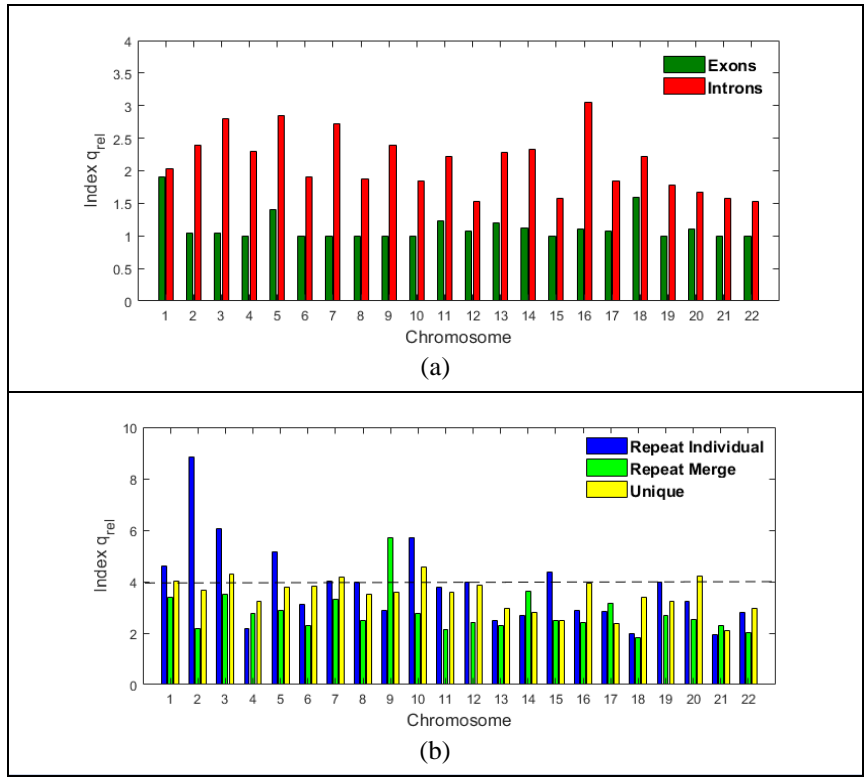

*Figure 6:* *The estimation of $q_{rel}$ index in per chromosome and genomic entity: **(a)** Exons, Introns, **(b)** Repeats, Unique*

Finally, as concern the $q_{sen}$ index (Figure 7), there is a strong differentiation between chromosomes in exons and introns region (Figure 7a). As one can observe, the $q_{sen}$ index in chromosomes within introns takes much higher values than those in exons. This mean that the multifractal character of the chromosomes is stronger within introns. The multifractal profile verifies the presence of different scaling in physical space which characterized the different order of information per region and per chromosome in all genome. Moreover, the multifractal character is different between chromosomes within the introns. Oppositely, in the exons the multifractal character has almost the same behavior for the most of chromosomes. Similar results to the exons, we observe for the repeat and unique genomic entities, but with smaller values of $q_{sen}$ index (Figure 7b). In certain chromosomes, there is no differentiation of multifractal character between different genomic entities. Moreover, there is a differentiation of multifractal character between chromosomes with higher values than the repeat and unique ones.

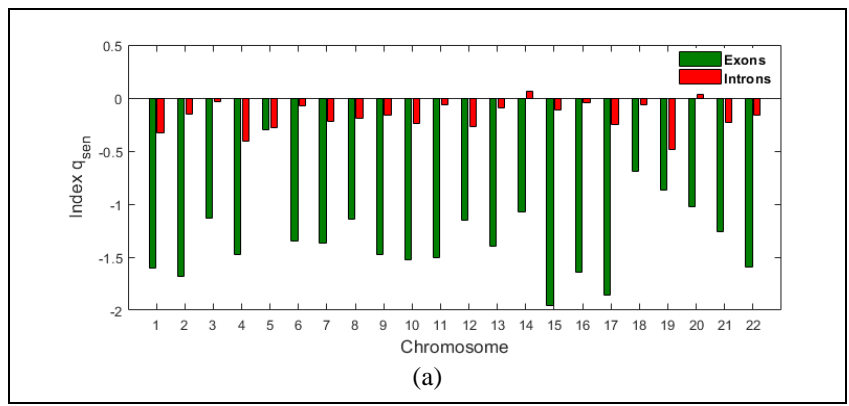

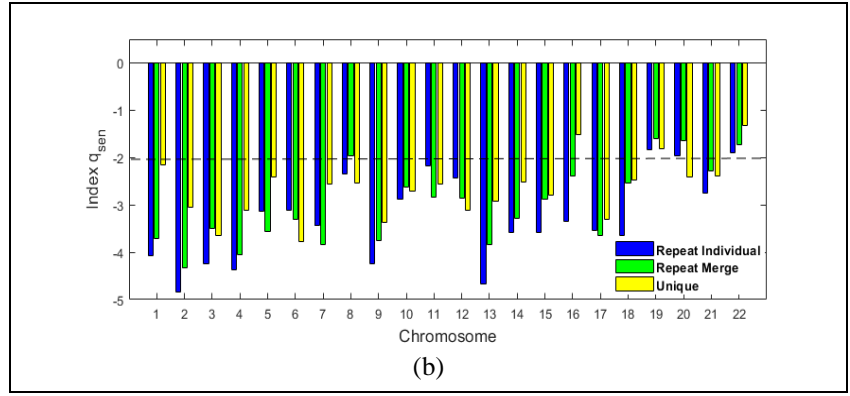

*Figure 7:* *The estimation of $q_{sen}$ index per chromosome genomic entities:* *(a)* *Exons, Introns,* *(b)* *Repeats, Unique.*

### 3.1.3 Correlation Dimension

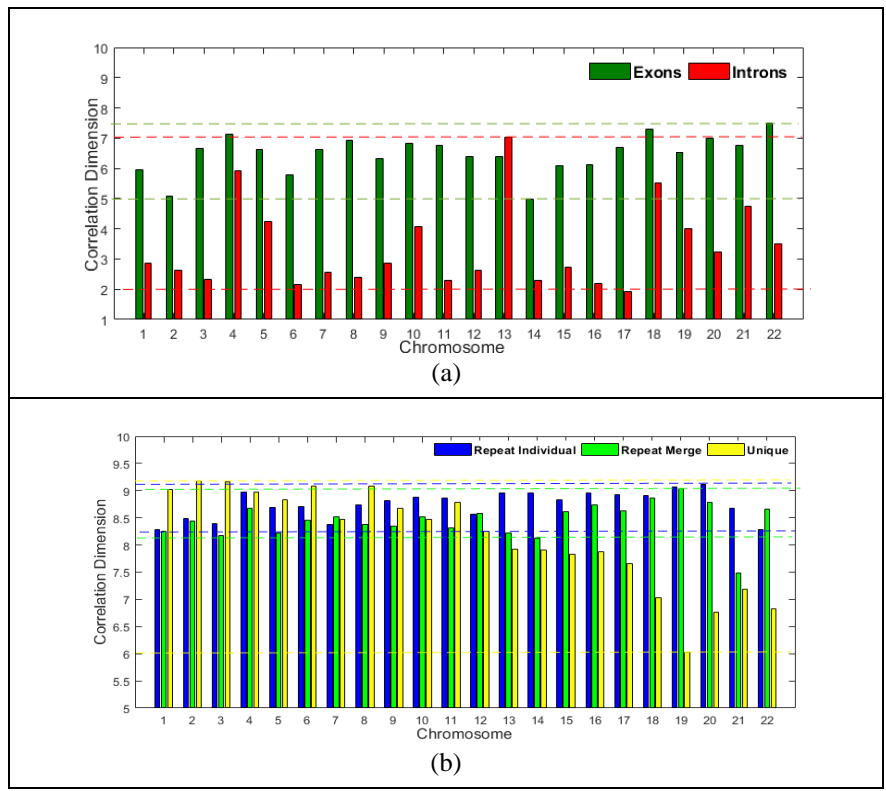

*Figure 8:* *The estimation of Correlation Dimension ($D_2$) per chromosome and genomic entity.* *(a)* *Exons, Introns,* *(b)* *Repeats, Unique.*

In Figure 8, the estimation of correlation dimension ($D_2$) is presented. For a random system the correlation dimension is approaching the embedding dimension. In contrast a more deterministic self-organized system the correlation dimension remains at lower prices from the embedding dimension. The estimation of the correlation dimension showed that the distribution of the sizes of the intronic regions reveals strong self-organization with strong variations per chromosome. The self-organized behavior means the existence of fundamental laws which produced the order of the sizes of intronic regions. Moreover, as we observe in Figure 8a, there is a differentiation between Chromosomes, but the important thing here is the reduction of dimensionality of Intron and Exon genomic entities and even more the significant reduction of dimensionality of Introns against Exons. As one can see, the correlation dimension for the Introns is $D_2 \leq 5$ for almost all Chromosomes (except Chromosomes 4, 13 and 18), while the correlation dimension for the Exons is $D \geq 5$ for all Chromosomes. In Figure 8b, we observe the correlation dimension for Repeat Individual, Repeat Merge and Unique signals and does not seem to be any differentiation between Chromosomes or genomic entities, except in cases of Chromosomes 18-22 where it is observed a significant reduction of dimensionality ($D \leq 7$) in Unique genomic entity and significant differentiation with the rest of Chromosomes.

## 3.2 Complexity Factor (COFA)

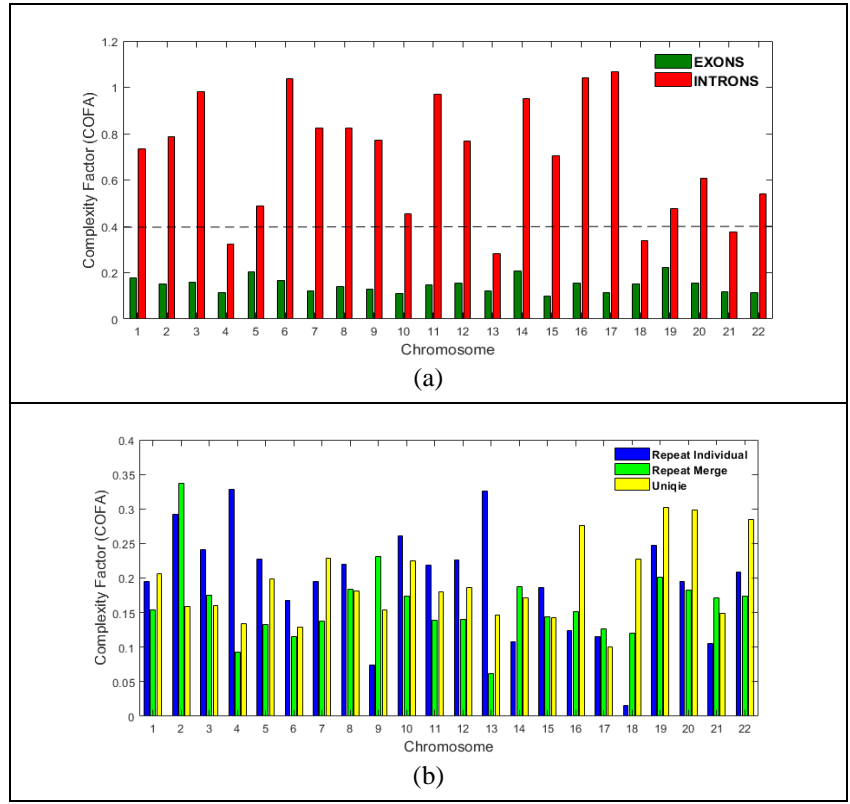

*Figure 9: The estimation of the technical term Complexity Factor (COFA) per chromosome and genomic entity: (a) Exons, Introns, (b) Repeat, Unique.*

In Figure 9, we presented the estimation of the technical term Complexity Factor (COFA) per chromosome and genomic entity is presented. As we observe in Figure 9a, there is a significant variation of COFA between genomic entities and Chromosomes. The Exons are characterized by a low complexity ($COFA < 0.2$), while the Introns by high complexity ($COFA > 0.6$). In Figure 9b, we observe Repeat Individual, Repeat Merge and Unique genomic entities, where characterized by low ($COFA < 0.2$) and medium ($0.2 < COFA < 0.6$) complexity.

## 3.3 Machine Learning Algorithms

In this section we used the estimation of complexity metrics as an input in ML algorithms for classification clustering and prediction with the thought to see if the variation of the metrics that correspond to each genomic entity for all chromosomes can be identified as a common dynamical feature which is characterizing these genomic entities. We analyzed these set of metrics with first a supervised classification based on Naïve Bayes classifier, and second with a k-means clustering.

### 3.3.1 Supervised Classification (Naive Bayes classifier)

We used the supervised classification based on Naïve Bayes classifier (see Appendix A for details). We prepare the model using, a different set of complexity metrics, every time we run the classification process. The Figure 10 shows the classification model's accuracy for each try. These tables, also known as Confusion Matrices, reveals true versus predicted values. The diagonal of each matrix represents the correct predictions. The first set of variables ($h, q_{stat}, q_{rel}, D_2, \Delta D_q$) gives the highest accuracy ((Correct predictions)/ (Number of Examples)) of 92,59%.

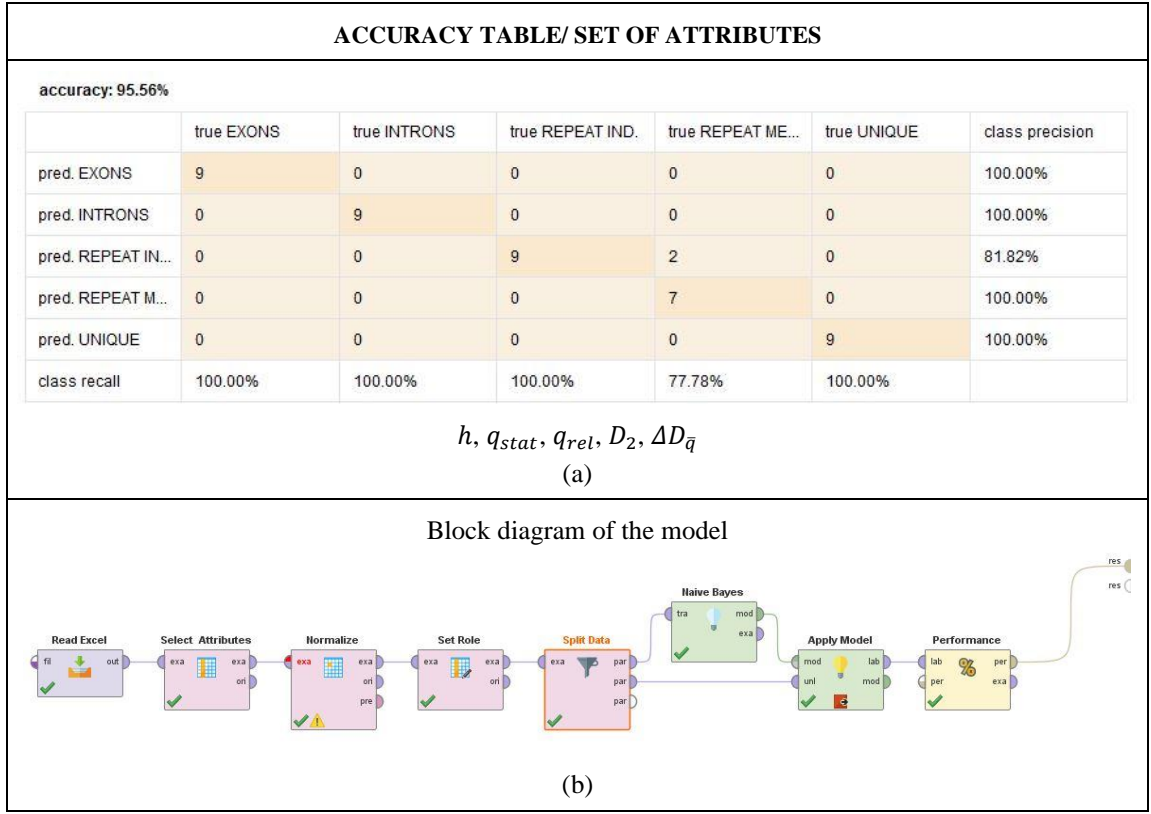

*Figure 10: (a) The classification model's accuracy for the best try, (b) The block diagram of the model.*

For the classifier's evaluation we used a 60/40 train/test set split. We split the dataset into a training dataset and a test dataset. Our model randomly selects 60% of the instances for training and use the remaining 40% as a test dataset. On the test dataset the accuracy of our model is:

- 95,56% with attributes: $h, q_{stat}, q_{rel}, D_2, \Delta D_{\bar{q}}$
- 95,56% with attributes: $h, q_{stat}, q_{sen}, q_{rel}, D_2, \Delta D_{\bar{q}}$
- 75,56% with attributes: $h, \sqrt{(q_{stat})^2 + (q_{rel})^2 + (q_{stat})^2}, D_2, \Delta D_{\bar{q}}$
- 77,78% with attributes: $COFA, \Delta D_{\bar{q}}$

*3.3.2 K-means Clustering (Unsupervised)*

Similarly, with sub-section 3.3.1, we applied the unsupervised k-means clustering (see Appendix A for details). We prepared the model using every time we run the clustering process a different set of complexity metrics. The best results are shown in Figure 11. Each bar shows the number of regions per chromosome which include to the cluster. It's clear that the model managed to separate the DNA regions in different clusters with very high accuracy on Exons, Introns and Unique and high accuracy on Repeat Individual. The region Repeat Merge had the least accuracy.

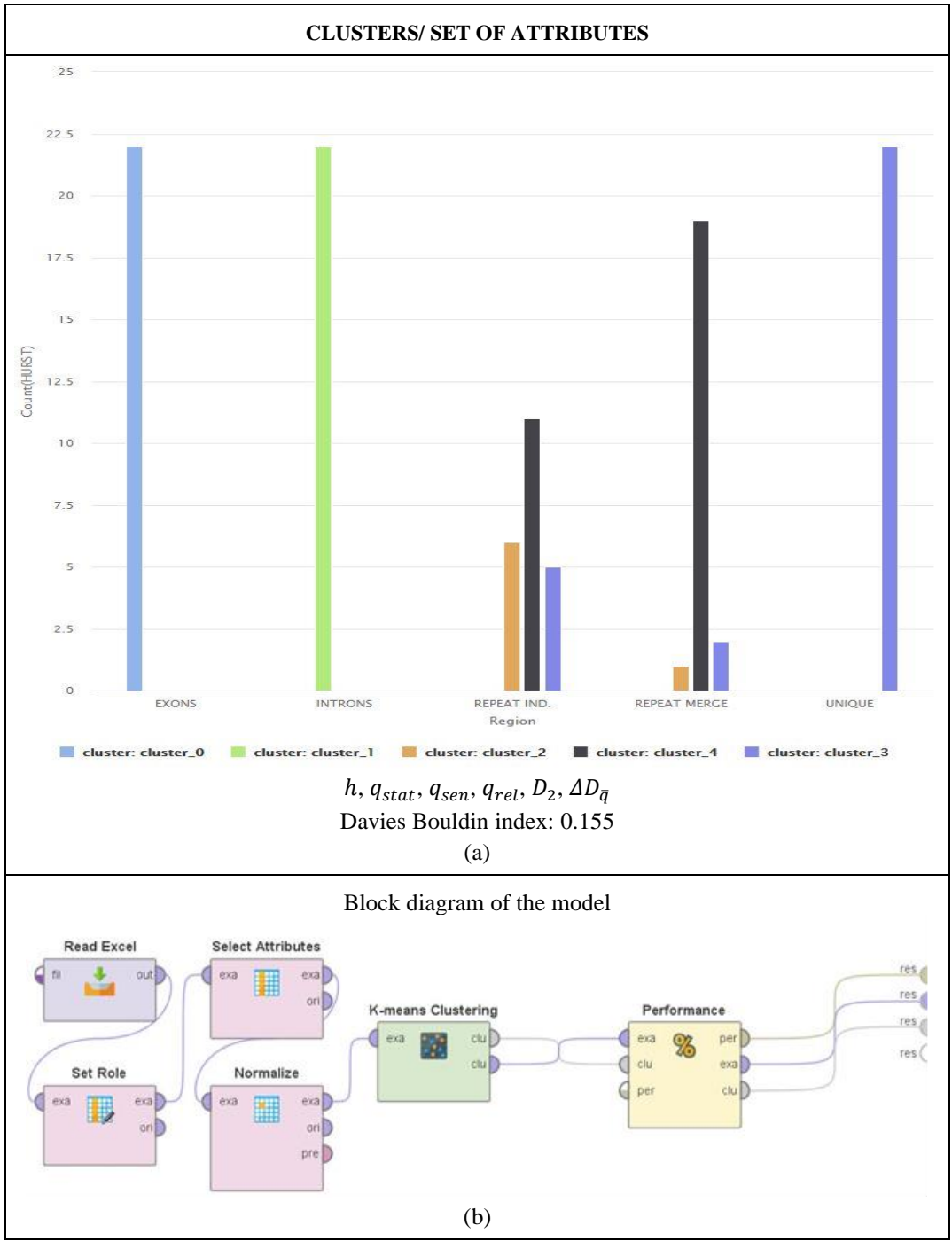

*Figure 11: (a) The clustering model's results for the best try; (b) The block diagram of the model.*

To evaluate each clustering process, we used the Davies-Bouldin (DB) index (Davies, 1979). The DB index provide an internal evaluation schema (the score is based on the cluster itself and not on external knowledge such as labels.) and is bounded from 0 to 1. Lower score is better.

In Figure 12 is presented the DB performance of each model with different number of attributes for different values of k parameter. The number, the type and the combination of attributes characterized the success of the model performance. The set of $h$, $q_{stat}$, $q_{sen}$, $q_{rel}$, $D_2$, $\Delta D_q$ complexity metrics gave the best DBI performance and specifically we had the lowest value for $k = 5$ parameter.

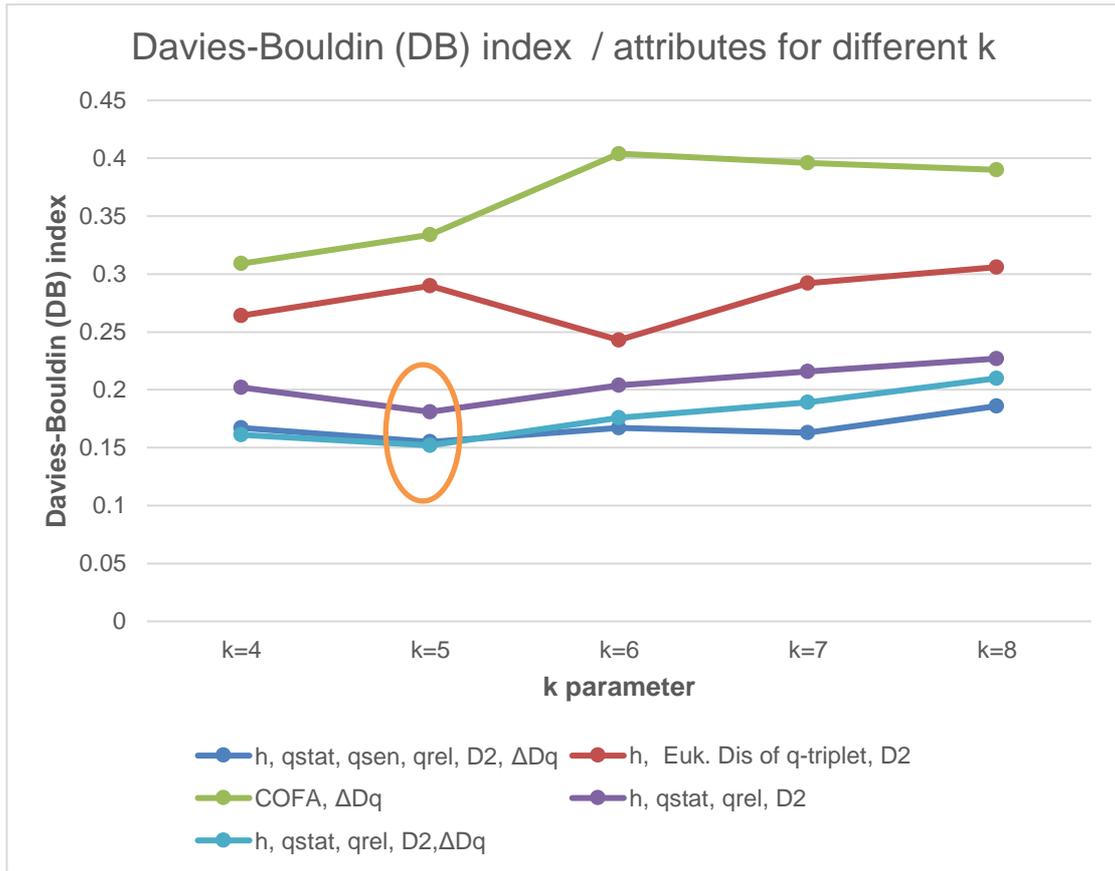

*Figure 12: The DB index performance versus number of attributes for different k.*

In Figure 13a, the 3d-scattered plot is presented with complexity metrics: h, qstat and D2. In Figure 13b, the results of k-means model, with the same complexity metrics, with clusters k=5 is shown. The variation of the complexity metrics is identified from the clustering model in very high accuracy for regions: Introns, Exons, Unique and high-medium accuracy for the rest of the regions.

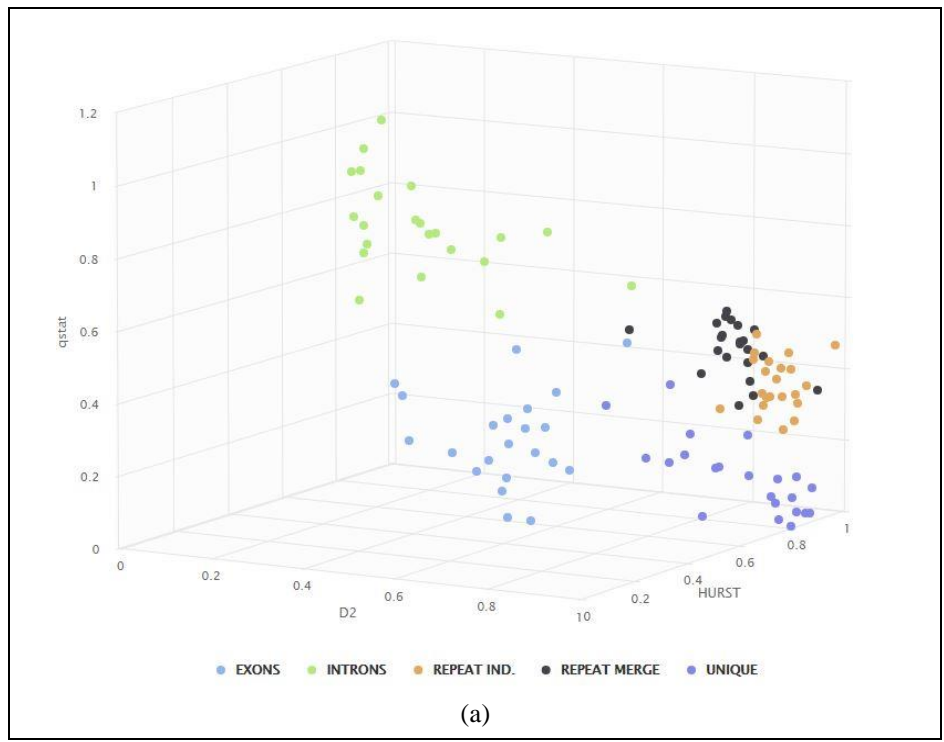

(a)

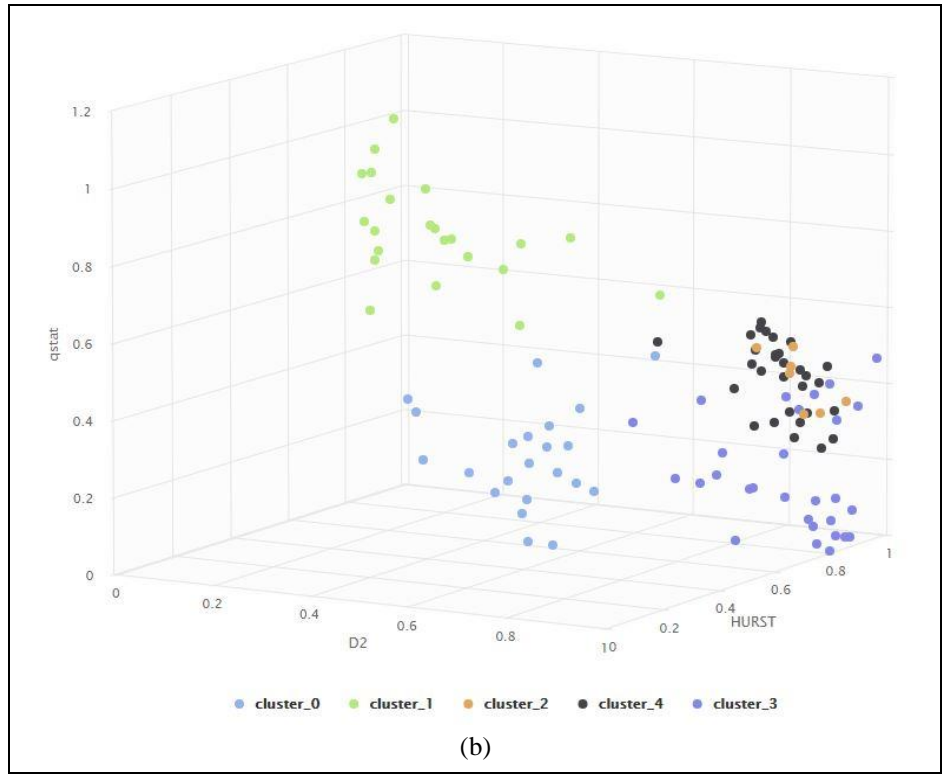

(b)

*Figure 13: Visualization of the regions clusters in all genome: (a) real 3d visualization, (b) 3d visualization after k-means clustering (k=6).*

*3.3.3 K-means Clustering (Unsupervised) based on COFA index*

Similarly, with sub-section 3.3.1 and 3.3.2, we applied the unsupervised k-means clustering (see Appendix A for details) based on the COFA metric for different values of parameter k. The best results for the DB index versus the k parameter we presented in Figure 14. For the parameters k=5 we had the lowest values of DB index performance. This mean that the k-means model creates 5 clusters.

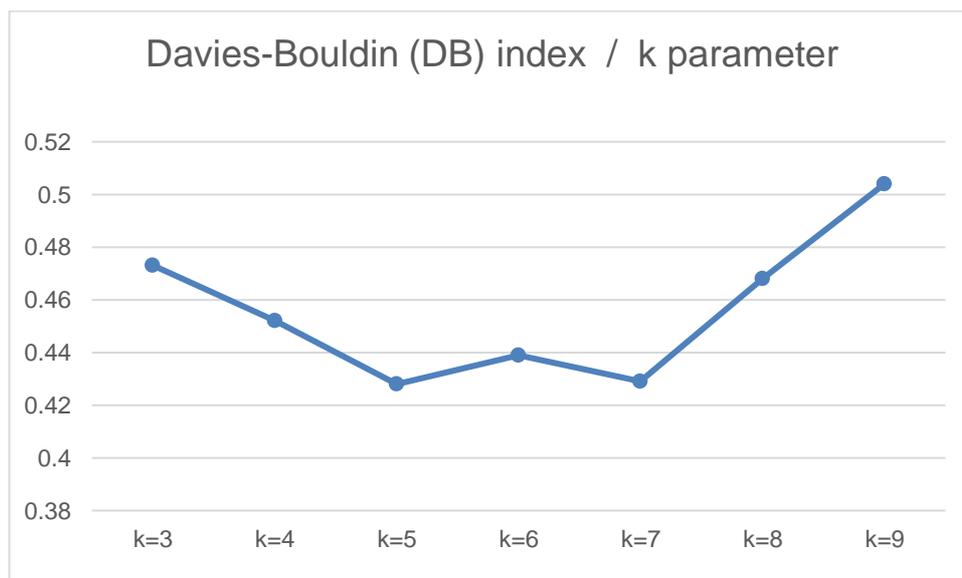

*Figure 14: The DB index performance versus parameter k (where k is the number of clusters).*

In Figure 15 the clusters for the best DB index performance are presented. In every cluster are included a set of different genomic entities from different chromosomes with a common geometrical center of the variations of the COFA index. With this method of clustering based on the COFA index we discriminated sets of genomic entities per chromosomes which appears similar dynamics or dynamics which lives around a local center. These sets may contain specific flows of information which are produced from fundamental laws and symmetries. It would be promising to see these findings in the lab.

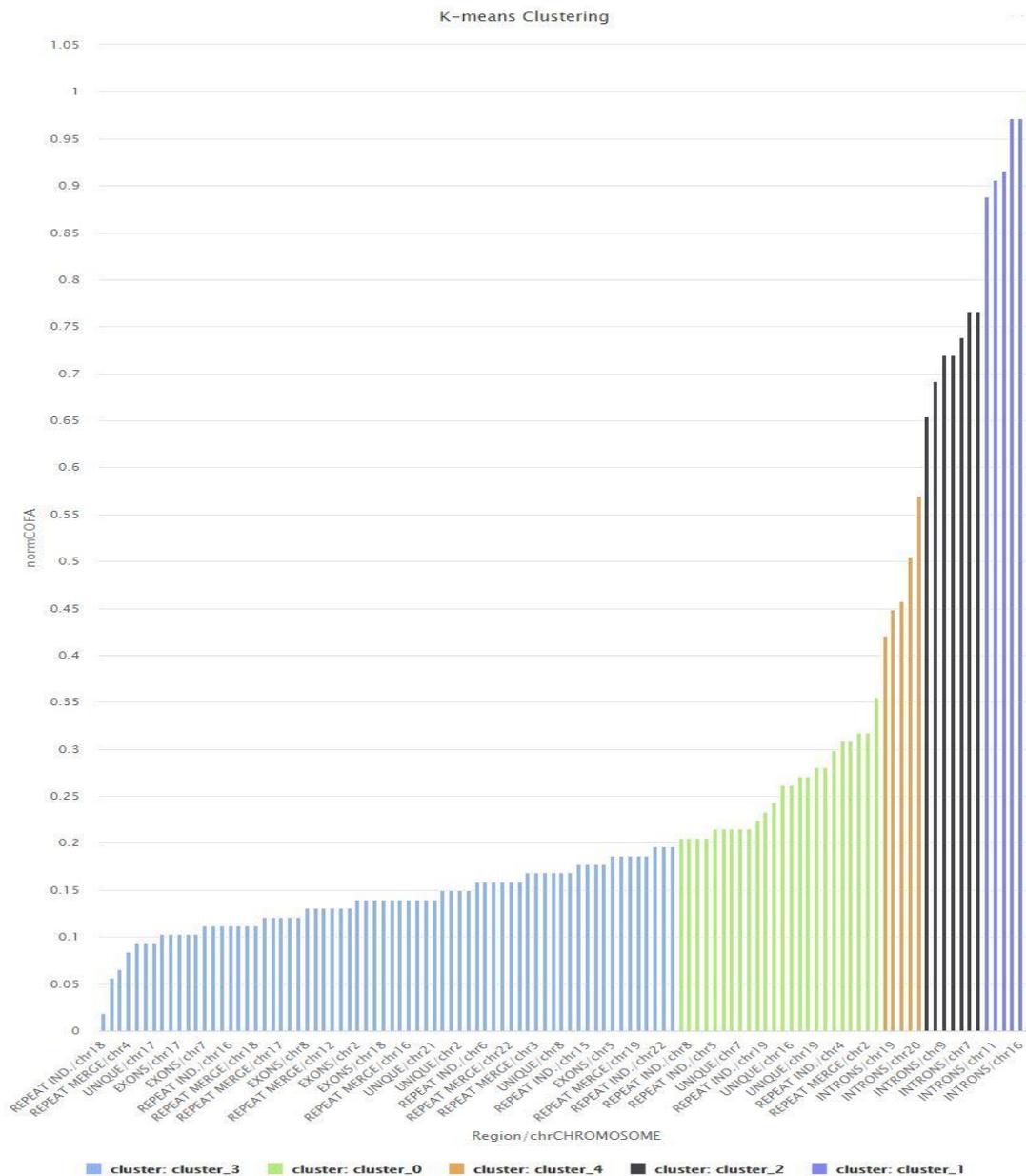

*Figure 15:* Visualization of the clusters in all genome after k-means clustering (k=5) based on COFA index. We separate the results in five clusters. In the x-axis are the genomic entities and chromosome reference, for example, the first one is: Repeat individual genomic entity in the chromosome 18, and the y-axis is the COFA index.

## 4. Discussion - Closing Remarks

In this study, the DNA distribution of regions of genome, were used to understand the degree of complexity behavior within exonic, intronic, repeat individual, repeat merge and unique regions of the genome. The analysis was based upon in complexity metrics to phase or physical space with the estimation of Hurst exponent, multifractal indices, q-triplet of Tsallis, correlation dimension and complexity factor (COFA) index and presented variations in the degree of complexity behavior per region and chromosomes. In particular, the low-dimensional deterministic nonlinear chaotic dynamics (anomalous random walk-strange dynamics) and the non-extensive statistical character of the DNA sequences was verified with strong multifractal characteristics and long-range correlations.

The results of this study reveal the DNA chromosomic system as a dynamical system working throughout an anomalous random walk and strange dynamical process underlying the biological temporal evolution and creating the DNA multifractal structure system. The multifractal DNA character reveals the DNA system as a globally unified, multiscale self-correlated and information storage of fractal system. The evolution of the chromosomic system includes consecutive critical points and self-organizing phase transition scenario included in the DNA dynamics. This process creates critical self-organized states with the DNA storage of information redundancy according to the DNA entropy reduction and self-organization. This process corresponds to the maximization of Tsallis entropy function at different chromosomic regions. The DNA chromosomic system includes scales and fundamental laws everywhere as the DNA entities are built throw the underlying DNA strange dynamics.

Moreover, the findings of this study reveal the chromosomic DNA system as the storage of biological and physical information of which small percentage has decoded. In this direction the complexity theory and the computational tools can lead to the further decoding of the DNA hidden information. In addition, the Tsallis theory were used in this study showed the existence of the non-Gaussian character everywhere in the DNA. The DNA system is a storage of meaningful biological information that remains to be decoded and to be related with new and unknown biological processes.

Notably the results of the hurst exponent reveals that the distributions of all regions in genome are characterized by memory character or persistent behavior in all chromosomes. Specifically, this memory character has very strong variations per chromosome in exonic and intronic region. The chromosome 13 in exonic region has anti-persistent behavior and the chromosomes 4,10,15,18 are almost equal to a Brownian (normal or fractional) Gaussian anomalous diffusion.

The results of the $q$ stationary reveals that the distribution of the regions in genome characterized by long range correlations. This non-extensive behavior is stronger in intronic region with strong variations per chromosome. Similarly, the variations are very strong in the exonic region. The lengths on the distributions of the regions is determined by previous lengths with an internal dynamical procedure.

The results of the $q$ relaxation suggest that the distribution of all regions may reach a new metastable state with different time profiles. Clearly though while all regions include information, they are of a complex character, such that there are differences in the degree of complexity and therefore this complexity impacts the time they take to transition to a new state of equilibrium upon being disturbed.

Similarly, the results of the q sensitivity reveal that the distribution of regions have strong multifractal profile in all genome and significant variations per chromosome. The multifractal profile verifies the presence of different scaling in physical space which characterized the different order of information per region and per chromosome in all genome.

In the reconstructive phase space, the distribution of the intronic region reveals strong self-organization with strong variations per chromosome. In chromosomes 1,2,3,6,7,8,11,12,14-17 we observed very strong self-organization against the other chromosomes of the genome of the aforementioned region. The same we observed in the chromosomes 1,2,6,17 of the regions exonic and in the chromosome 19 of unique region. The above results indicate the presence of nonlinear and low-dimensional DNA dynamics underlying the DNA distribution of regions.

The new technical index complexity factor (COFA) which represents the geometrical measure of the complexity in a Euclidean space successfully used as a new innovative external classifier with important findings of information islands based on ML models. The COFA index is based on the hurst exponent, the Euclidean distance of the $q$−triplet of Tsallis statistics and the correlation dimension. There is an open scientific research field for the modifications of the COFA index with linear or non-linear modes and with other sets of complexity metrics. That will be presented in other future studies.

The results showed that the underlying dynamical processes, which give rise to the organization of the genome, correspond to the extremization of $q$ − entropy principle included in the non-extensive statistical mechanics of Tsallis (Tsallis, 1988; 2002). The $q$ − entropy principle of Tsallis applies the unification of the macroscopic to the microscopic level through the multiscale interaction and the scale invariance principle included in the power laws of complex phenomena.

The projection of the dynamics to the statistics in the phase space develops a complete picture which integrated to the variations of the complexity metrics. The redundancy of information in DNA lies between randomness and order in a continuous evolutionary process of thousands of years (Beltrami, 1999) based on fluctuations and deterministic laws. This picture of dynamics can be identified from machine learning tools for clustering, classification and prediction. The results of the ML tools successfully identified the different degree of complexity profile of the distribution of the regions with high accuracy, based on a given set of complexity metrics. In conclusion, the distribution of the size of the genome entities is characterized from different degree of complexity profiles, which is recognizable from the machine learning models. This integrated methodology (Fig. 3) consisted a new innovative methodology for the identification of the symmetries and fundamental laws which produces the order of information in all genome and generate strange dynamics which is observable and qualitatively measurable. This picture of strange dynamics is identified from machine learning (ML) tools for clustering, classification and prediction with high accuracy.

Finally, the merging of interdisciplinary of complexity theory and genetics can provide semantic results in the direction of a deeper understanding and promotion of the fundamental laws of biology with new motifs, patterns and interactions of the complex biological information.

**Author Contributions:** conceptualization, L.P.K., G.P.P., D.S.M. ; methodology, L.P.K.; software, L.P.K., E.G.P., G.T.; formal analysis, L.P.K., E.G.P., G.T.; investigation, T.L.M., J.L.D.; resources, L.P.K., E.G.P., G.T.; data curation, T.L.M., J.L.D.; writing-original draft preparation, L.P.K., E.G.P., G.T., G.L.S., G.P.P., D.S.M. ; writing-review and editing, L.P.K., E.G.P., G.T., G.L.S., G.P.P., D.S.M.; visualization, L.P.K., E.G.P., G.T.; supervision, L.P.K., D.S.M.; project administration, L.P.K., D.S.M.;

**Funding:** This research was partially funded by PENN, grant number 27115-260850000

**Conflicts of Interest:** The author declares no conflict of interest.

## Appendix A. Supplementary Information

### A.1. Theoretical Framework

The DNA chromosomic system taking into account the nonlinear and strange dynamics can be described from the general equation:

$$\frac{d\vec{X}(\vec{r},t)}{dt} = F_\lambda(\vec{X}, W) \qquad (2)$$

where the vector $\vec{X}$ describes the state of the chromosomic chemical system, while the nonlinear function $F(\vec{X}, W)$ describes the temporal change $d\vec{X}/dt$ of the state vector. The state vector evolves temporarily in the state space of the biological evolution process. The control parameter $\lambda$ describes the degree of physical connection of the DNA system with its biological and chemical environment while the quantity $W$ corresponds to the temporal evolution of the system connection with its environment. The environment state function $W$ can be high or low dimensional. The dimension of the DNA state vector $\vec{X}$ and the topology of the corespondent DNA state space can change according to the control parameter values. As the control parameter $\lambda$ changes the profile of the dynamics of the system change also through phase transition self-organization of the entire system. Complexity theory is related with the nonlinear and strange dynamics included to the equation (2) and the statistical character of the system evolution in the state space. The multifractal topology of the state space is created through the entropy maximization principle (Pavlos, 2015; Karakatsanis, 2018).

### A.1.1 Non-extensive statistical mechanics

The non-extensive statistical theory is based mathematically on the nonlinear equation:

$$\frac{dy}{dx} = y^q, (y(0) = 1, q \in R \qquad (3)$$

with solution the $q-$ exponential function such as: $e_q^x = [1 + (1-q)x]^{\frac{1}{1-q}}$. For further characterizing the non-Gaussian character of the dynamics, we proceed to the estimation of Tsallis $q-$ triplet based on Tsallis nonextensive statistical mechanics. Nonextensive statistical mechanics includes the $q-$ analog (extensions) of the classical Central Limit Theorem (CLT) and $\alpha-$ stable distributions corresponding to dynamical statistics of globally correlated systems. The $q$ − extension of CLT leads to the definition of statistical $q$ −parameters of which the most significant is the $q$ − triplet($q_{sen}, q_{rel}, q_{stat}$), where the abbreviations $sen, rel$, and $stat$, stand for sensitivity (to the initial conditions), relaxation and stationary (state) in nonextensive statistics respectively (Tsallis, 2004; 2011; Umarov, 2008). These quantities characterize three physical processes: a) $q$ − entropy production ($q_{sen}$), (b) relaxation process ($q_{rel}$), c) equilibrium fluctuations ($q_{stat}$). The $q$ − triplet values characterize the attractor set of the dynamics in the phase space of the dynamics and they can change when the dynamics of the system is attracted to another attractor set of the phase space. Equation (3) for $q = 1$ corresponds to the case of equilibrium Gaussian (Boltzmann-Gibbs (BG)) world [34]. In this case, the $q$ − triplet of Tsallis simplifies to $q_{sen} = 1, q_{stat} = 1, q_{rel} = 1$.

### A.1.2 $q$ −triplet of Tsallis theory ($q_{sen}, q_{rel}, q_{stat}$).

*a) $q_{stat}$ index*

A long-range-correlated meta-equilibrium non-extensive process can be described by the nonlinear differential equation (Tsallis, 2009):

$$\frac{d(p_i Z_{q_{stat}})}{dE_i} = -\beta q_{stat}(p_i Z_{q_{stat}})^{q_{stat}} \tag{4}$$

The solution of this equation corresponds to the probability distribution:

$$p_i = \frac{e_{q_{stat}}^{-\beta_{stat} E_i}}{Z_{q_{stat}}} \tag{5}$$

where $\beta_{q_{stat}} = \frac{1}{KT_{stat}}$, and $Z_{q_{stat}} = \sum_j e_{q_{stat}}^{-\beta q_{stat} E_j}$. Then the probability distribution is given:

$$p_i \propto [1 - (1-q)\beta_{q_{stat}} E_i]^{1/1-q_{stat}} \tag{6}$$

for discrete energy states $\{E_i\}$ and by

$$p(x) \propto [1 - (1-q)\beta_{q_{stat}} x^2]^{1/1-q_{stat}} \tag{7}$$

for continuous $x$ states of $\{X\}$, where the values of the magnitude $X$ correspond to the state points of the phase space. Distributions functions (6) and (7) correspond to the attracting stationary solution of the extended (anomalous) diffusion equation related to the nonlinear dynamics of the system. The stationary solutions $P(x)$ describe the probabilistic character of the dynamics on the attractor set of the phase space. The non-equilibrium dynamics can evolve on distinct attractor sets, depending upon the control parameters, while the $q_{stat}$ exponent can change as the attractor set of the dynamics change. For the estimation of Tsallis $q$ − Gaussian distributions we use the method described in Ferry et al. (Ferry, 2010).

*b) $q_{sen}$ index*

Entropy production is related to the general profile of the attractor set of the dynamics. The profile of the attractor can be described by its multi-fractality as well as by its sensitivity to initial conditions. The sensitivity to initial conditions can be expressed as:

$$\frac{d\xi}{dt} = \lambda_1 \xi + (\lambda_q - \lambda_1)\xi^q \tag{8}$$

where $\xi$ is the trajectory deviation in the phase space: $\xi \equiv \lim_{\Delta(x)\to 0}\{\Delta x(t)\backslash\Delta x(0)\}$ and $\Delta x(t)$ is the distance between neighbouring trajectories (Tsallis, 2002). The solution of equation (8) is given by:

$$\xi = \left[1 - \frac{\lambda_{q_{sen}}}{\lambda_1} + \frac{\lambda_{q_{sen}}}{\lambda_1}e^{(1-q_{sen})\lambda_1 t}\right]^{\frac{1}{1-q_{sen}}} \tag{9}$$

The $q_{sen}$ exponent is related to the multi-fractal profile of the attractor set according to

$$\frac{1}{q_{sen}} = \frac{1}{\alpha_{min}} - \frac{1}{\alpha_{max}} \tag{10}$$

where $\alpha_{min}, \alpha_{max}$ corresponds to zero points of the multi-fractal exponent spectrum $f(\alpha)$, that is $f(\alpha_{min}) = f(\alpha_{max}) = 0$. For the estimation of the multifractal spectrum we use the method described in Pavlos et al. (Pavlos, 2014).

By using $D_{\bar{q}}$ spectrum we estimate the singularity spectrum $f(\alpha)$ using the Legendre transformation:

$$f(\alpha) = \bar{q}a - (\bar{q} - 1)D_{\bar{q}} \tag{11}$$

where $\alpha = \frac{d\tau(\bar{q})}{d\bar{q}}$. We note that the Tsallis q-entropy number is a special number corresponding to the extremization of Tsallis entropy of the system, while the $\bar{q}$ describe the range of real values of generalized dimension spectrum $D_{\bar{q}}$.

The degree of multifractality is given by:

$$\Delta a = a_{max} - a_{min} \tag{12}$$

and the degree of asymmetry $A$ can be estimated by the relation:

$$A = \frac{a_0 - a_{min}}{a_{max} - a_0} \tag{13}$$

In particular, $\alpha_0$ corresponds to the largest fractal dimension, which in this case is $f(\alpha) = 1$. It is important to note here that the singularity exponents $\alpha$ of the singularity spectrum $f(\alpha)$ corresponds to the Holder exponent and reveal the intensity of the topological singularity of the phase space as well as how irregular are the physical magnitudes defined in the phase space of the system. The value $\alpha_0$, separates the values of $\alpha$ in two distinct intervals, $\alpha < \alpha_0$ and $\alpha > \alpha_0$ with different physical meaning. In particular, the left part of singularity spectrum $f(\alpha)$ is related with values $\alpha$ lower than the value $\alpha_0$, and correspond to the low dimensional regions of the phase space, which is described by the right part of $D_{\bar{q}}$ spectrum. Similarly, the right part of the singularity spectrum $f(\alpha)$ is related with values $\alpha$ higher than $\alpha_0$ and correspond to the high dimensional regions of the phase space, which is described by the left part of the curve $D_{\bar{q}}$ of the generalized dimension spectrum.

According to these characteristics of $f(\alpha)$ and $D_{\bar{q}}$ spectra, the high dimensional regions of phase space includes smoother fractal topology than the low dimensional regions, where the fractal character is stronger. Low dimensional regions of phase space cause strong fractional acceleration and anomalous diffusion processes of the experimental TMS. The estimation of $\Delta D_{\bar{q}}$ between the low ($\bar{q} \to +\infty$) and high ($\bar{q} \to -\infty$) dimensional regions of the phase space reveals the multifractal behavior of the system. High (Low) values of $\Delta D_{\bar{q}}$ shows strong (weak) multifractality.

*c) $q_{rel}$ index*

Thermodynamic fluctuation – dissipation theory is based on the Einstein original diffusion theory (Brownian motion theory). Diffusion is a physical mechanism for extremization of entropy. If $\Delta S$ denote the deviation of entropy from its equilibrium value $S_0$, then the probability of a proposed fluctuation is given by:

$$P \approx \exp\left(\frac{DS}{k}\right) \tag{14}$$

The Einstein – Smoluchowski theory of Brownian motion was extended to the general Fokker Planck (FP) diffusion theory of non-equilibrium processes. The potential of FP equation may include many meta-equilibrium stationary states near or far away from thermodynamical equilibrium. Macroscopically, relaxation to the equilibrium stationary state of some dynamical observable $O(t)$ related to system evolution in the phase space can be described by the form of general form:

$$\frac{d\Omega}{dt} \cong -\frac{1}{\tau}\Omega \tag{15}$$

Where $\Omega(t) \equiv [O(t) - O(\infty)]/[O(0) - O(\infty)]$ describes the relaxation of the macroscopic observable $O(t)$ towards its stationary state value. The non-extensive generalization of fluctuation – dissipation theory is related to the general correlated anomalous diffusion processes (Tsallis, 2009). The equilibrium relaxation process is transformed to the meta-equilibrium non-extensive relaxation process according to:

$$\frac{d\Omega}{dt} = -\frac{1}{T_{q_{rel}}} \Omega^{q_{rel}} \qquad (16)$$

the solution of this equation is given by:

$$\Omega(t) \cong e_{q_{rel}}^{-t/\tau_{rel}} \qquad (17)$$

The autocorrelation function $C(t)$ or the mutual information $I(t)$ can be used as candidate observables $\Omega(t)$ for estimation of $q_{rel}$. However, in contrast to the linear profile of the correlation function, the mutual information includes the nonlinearity of the underlying dynamics and it is proposed as a more faithful index of the relaxation process and the estimation of the Tsallis exponent $q_{rel}$.

### A.1.3 Correlation Dimension ($D_2$)

In the reconstructed phase space, we estimate the correlation integral $C(R, m)$ as function of radius $R$ and embedding dimension $m$ with parameter $w$ of Theiler according to the relation:

$$C_2(r) = \frac{2}{(N-W)(N-W-1)} \sum_{i=1}^{N} \sum_{j=1+W+1}^{N} \Theta(r - \|x(i) - x(j)\|) \qquad (18)$$

When the dynamics is nonlinear then it is possible in the phase space to exist low-dimensional attracting sets (strange attractors). In this case the correlation integral reveals power law scaling profile: $C(R) \sim R^D$ where $D$ is the mean fractal dimension of the strange attractor. According to Taken's (Takens, 1981) the reconstructed m-dimensional state space is an efficient embedding for $m \geq 2D + 1$. Efficient embedding means topological equivalence between the real and the reconstructed state space (Broomhead, 1986). According to these theoretical concepts when there exist a low-dimensional chaotic (strange) attractor of the dynamics then $D$ takes fractal value and corresponds to the saturation value of the slopes $d_m$ of the correlation integrals as embedding dimension increases according to the relation: $D = \lim_{m \to \infty} d_m$. The values $d_m$ are the scaling exponents of the correlation integral $C(R, m)$ for low values of $r(R \to 0)$ according to the relation:

$$C(R, m) \sim R^{d_m}, d_m = \frac{d[lnC(R,m)]}{d[\ln(R)]} \text{ for } R \to 0 \qquad (19)$$

### A.1.4 Hurst Exponent ($h$)

The Hurst exponent ($h$) related to the fractal dimension ($D$). The relationship between the fractal dimension and the Hurst exponent is:

$$D = 2 - h \qquad (20)$$

The fractal dimension shows how rough a surface is. A small value of Hurst exponent shows a higher fractal dimension and a rougher surface. A larger Hurst exponent shows a smaller fractional dimension and a smoother surface.
The values of the Hurst exponent range between 0 and 1. A value of 0.5 indicates a true random process (a Brownian time series). A Hurst exponent value $h, 0.5 < h < 1$ indicates "persistent behavior". Here an increase (decrease) probably followed by an increase (decrease). A Hurst exponent value $0 < h < 0.5$ indicates "anti-persistent behavior". Here an increase (decrease) probably followed by a decrease (increase). For the estimation of the Hurst exponent ($h$) in this study we use Rescaled Range Analysis (R/S) (Weron, 2002).
The Hurst exponent ($h$), is defined in terms of the asymptotic behavior of the rescaled range (R/S) as a function of the time span of a time series as follows:

$$E\left[\frac{R(n)}{S(n)}\right] = Cn^h, n \to \infty \qquad (21)$$

where, $R(n)$ is the range of the first $n$ values and $S(n)$ is their standard deviation, $E(x)$ is the expected value, $n$ is a number of data points in a time series and $C$ is a constant.

### A.1.5 Machine Learning Analysis

*a) Clustering*

Cluster analysis or clustering is the task of grouping a set of objects in such a way that objects in the same group (called a cluster) are more similar (in some sense) to each other than to those in other groups (clusters). It is a main task of exploratory data mining, and a common technique for statistical data analysis, used in many fields, including machine learning, pattern recognition, image analysis, information retrieval, bioinformatics, data compression, and computer graphics.

*b) k-means model*

k-means clustering is one of the simplest and popular unsupervised machine learning algorithms. Is a method of vector quantization, originally from signal processing. k-means clustering aims to partition n observations (Examples) into k clusters in which each observation belongs to the cluster with the nearest mean, serving as a prototype of the cluster. Clustering can be used on unlabeled data.

**How the k-means algorithm works**

The k-means algorithm determines a set of k clusters and assigns each Examples to exact one cluster. The clusters consist of similar Examples. The similarity between Examples is based on a distance measure between them. A cluster in the k-means algorithm is determined by the position of the center in the n-dimensional space of the n Attributes of the Example Set. This position is called centroid. It can, but do not have to be the position of an Example of the Example Sets. The k-means algorithm starts with k points which are treated as the centroid of k potential clusters. All Examples are assigned to their nearest cluster (nearest is defined by the measure type). Next the centroids of the clusters are recalculated by averaging over all Examples of one cluster. The previous steps are repeated for the new centroids until the centroids no longer move or max optimization steps is reached. The procedure is repeated max runs times with each time a different set of start points. The set of clusters is delivered which has the minimal sum of squared distances of all examples to their corresponding centroids.

The objective function for the k-means clustering algorithm is the squared error function:

$$J = \sum_{i=1}^{k} \sum_{j=1}^{n} (\|x_i - u_j\|)^2 = 1 \qquad (22)$$

where $\|x_i - u_j\|$ is the Euclidean distance between a point, $x_i$ and a centroid, $u_j$, iterated over all $k$ points in the $i^{th}$ cluster, for all $n$ clusters.

In simpler terms, the objective function attempts to pick centroids that minimize the distance to all points belonging to its respective cluster so that the centroids are more symbolic of the surrounding cluster of data points. K-means clustering is a fast, robust, and simple algorithm that gives reliable results when data sets are distinct or well separated from each other in a linear fashion. It is important to keep in mind that k-means clustering may not perform well if it contains heavily overlapping data, if the Euclidean distance does not measure the underlying factors well, or if the data is noisy or full of outliers.

*c) Supervised classification (Naive Bayes classifier)*

For this work we used the *Naive Bayes classifier* for the classification process.

Naive Bayes is a high-bias, low-variance classifier, and it can build a good model even with a small data set. It is simple to use and computationally inexpensive. Typical use cases involve text categorization, including spam detection, sentiment analysis, and recommender systems. The Naive Bayes Classifier technique is based on the so-called Bayesian theorem and is particularly suited when the dimensionality of the inputs is high. Despite its simplicity, Naive Bayes can often outperform more sophisticated classification methods.

Naive Bayes classifiers can handle an arbitrary number of independent variables whether continuous or categorical. Given a set of variables, $X = \{x_1, x_2, \dots, x_d\}$, we want to construct the posterior probability for the event $C_j$ among a set of possible outcomes $C = \{c_1, c_2, \dots, c_d\}$. In a more familiar language, $X$ is the predictors and $C$ is the set of categorical levels present in the dependent variable. Using Bayes' rule:

$$p(C_j \vee x_1, x_2, \dots, x_d) \propto p(x_1, x_2, \dots, x_d \vee C_j) p(C_j) \qquad (23)$$

where $p(C_j \vee x_1, x_2, ..., x_d)$ is the posterior probability of class membership, i.e., the probability that $X$ belongs to $C_j$. Since Naive Bayes assumes that the conditional probabilities of the independent variables are statistically independent, we can decompose the likelihood to a product of terms:

$$p(X \vee C_j) \propto \prod_{k=1}^{d} p(x_k \vee C_j) \quad (24)$$

and rewrite the posterior as:

$$p(C_j \vee X) \propto p(C_j) \prod_{k=1}^{d} p(x_k \vee C_j) \quad (25)$$

Using Bayes' rule above, we label a new case $X$ with a class level $C_j$ that achieves the highest posterior probability. Although the assumption that the predictor (independent) variables are independent is not always accurate, it does simplify the classification task dramatically, since it allows the class conditional densities $p(x_k \vee C_j)$ to be calculated separately for each variable, i.e., it reduces a multidimensional task to a number of one-dimensional ones. In effect, Naive Bayes reduces a high-dimensional density estimation task to a one-dimensional kernel density estimation. Furthermore, the assumption does not seem to greatly affect the posterior probabilities, especially in regions near decision boundaries, thus, leaving the classification task unaffected.

**Evaluation - Split Test**

For the classifier's evaluation we used a 60/40 train/test set split. The split of the dataset is a simple way to use one dataset to both train and estimate the performance of the classifier. We split the dataset into a training dataset and a test dataset. Our model randomly selects 60% of the instances for training and use the remaining 40% as a test dataset.